%% file: main.tex
\documentclass[fleqn,usenatbib]{mnras} 
\usepackage{newtxtext,newtxmath}

\usepackage[T1]{fontenc}

\DeclareRobustCommand{\VAN}[3]{#2}
\let\VANthebibliography\thebibliography
\def\thebibliography{\DeclareRobustCommand{\VAN}[3]{##3}\VANthebibliography}

\usepackage{graphicx}	
\usepackage{amsmath}	
\usepackage{siunitx}
\usepackage{siunitx}
\newcommand{\dotdeg}{\rlap{.}^\circ}
\newcommand{\instrument}[1]{\textit{#1}}
\newcommand{\software}[1]{\textsc{#1}}
\newcommand{\red}[1]{#1} 
\setcitestyle{notesep={; }} 

\title[Wising up to CatWISE]{
    Wising up to CatWISE:
    using \red{simulation-based inference} to interpret the ecliptic bias 
    and confirm the cosmic dipole excess
}

\author[Oayda \& Lewis]{
    Oliver T. Oayda$^{1}$\thanks{E-mail: oliver.oayda@sydney.edu.au}
    and Geraint F. Lewis$^{2}$
    \\
    $^{1}$Sydney Institute for Astronomy, School of Physics A28, The University of Sydney, NSW 2006, Australia
}

\date{Accepted XXX. Received YYY; in original form ZZZ}

\pubyear{\the\year{}}

\begin{document}
\label{firstpage}
\pagerange{\pageref{firstpage}--\pageref{lastpage}}
\maketitle

\begin{abstract}
  We apply Simulation-Based Inference (`SBI') to the cosmic dipole problem
  for the first time,
  measuring the distribution of quasar counts over the sky in the
  CatWISE2020 (`CatWISE') sample.
  We show that the quadrupole anisotropy in CatWISE can be attributed
  to the correlation between \textit{WISE}'s scanning law and
  photometric uncertainty in the $W1$ and $W2$ magnitudes,
  inducing an Eddington bias which varies with sky position.
  After explicitly modelling this with SBI,
  we use a neural likelihood estimator to find the posterior distribution
  for CatWISE's dipole,
  confirming the presence of a dipole twice as large as the CMB expectation
  but more seriously misaligned with the CMB direction ($\approx 3 \sigma$).
  We also use our learned likelihood to infer the Bayesian evidence,
  learning that models which increase the scale of CatWISE's photometric errors
  are most favoured.
  This is strong evidence that the sample's errors are underestimated
  or that there is an additional, unresolved systematic producing the same effect 
  as Eddington bias.
  While our results indicate that the cosmic dipole excess
  is a persistent issue for $\Lambda$CDM,
  we showcase that SBI can untangle the subtle and complex
  systematic issues affecting any sample derived from real astronomical data.
\end{abstract}

\begin{keywords}
cosmology: observations --- cosmic background radiation --- quasars: general ---
methods: statistical
\end{keywords}



\section{Introduction}
The cosmic dipole excess is another serious challenge
to the cosmological principle \citep[`CP';][]{di-valentino}.
If the Universe is isotropic and homogeneous,
as assumed by the CP underpinning the 
Friedmann-Lema\^{i}tre-Robertson-Walker (`FLRW') metric,
then an observer positioned in the cosmic rest frame (`CRF')
should perceive an isotropic and homogeneous distribution of matter.
Since we see a temperature anisotropy in the Cosmic Microwave Background (`CMB'),
we conclude that we must be travelling with respect to the CRF at
$v_{\text{CMB}} = 369.82\pm0.11\,\text{km}\,\text{s}^{-1}$
towards $(l,b) = (264\dotdeg021, 48\dotdeg253)$ 
in Galactic coordinates \citep{planck2020}.
This motion should imprint an analogous dipole
in the distribution of cosmological sources \citep{ellis1984},
which we refer to as the `cosmic dipole'.
However, recent studies measure a dipole in quasars and radio galaxies
that is roughly two to three times as large as expected
\citep[for a recent review, see e.g.][]{secrest+25}.
If the cosmic dipole is genuinely inconsistent with the CMB dipole,
it means we must rethink its interpretation as due to our motion,
and more fundamentally rethink the assumption of the cosmological principle.

Measuring the cosmic dipole, however, is not straightforward.
The implicit assumption in the test of \citet{ellis1984} is that measurements
are identically performed over the celestial sphere.
In practice, any survey --- whether ground or space based ---
will have instrumental systematics that can impact the probability of source detection
at different points over the sky.
Since an instrumental difference in the rate of source detection will diminish or elevate
source counts,
this can generate a dipole and potentially high-order moments in the data,
biasing measurement of the cosmic dipole.
Accounting for these effects is therefore paramount.

Recently, \citet[][`S21']{secrest2021} and \citet[][`S22']{secrest2022}
measured the cosmic dipole in the CatWISE2020 (`CatWISE') sample of quasars
\citep{marocco2021}.
Despite masking out 30$^\circ$ above and below the Galactic plane and correcting for
dust extinction, the sample exhibits a prominent `ecliptic bias'.
Namely, counts of quasars are diminished near the ecliptic poles
and elevated around the ecliptic equator.
To mitigate its effect, the authors included an ecliptic correction,
\red{using a linear fit to density versus declination as a weighting function.}
While the CatWISE samples are slightly different across S21 and S22,
they report a similar discrepancy with the CMB dipole,
reaching a significance of $\approx 5\sigma$.

Since CatWISE is a key piece of evidence in the cosmic dipole anomaly,
it is important to understand its peculiarities.
This is especially relevant given that the cosmic dipole imprints a
subtle variation in source density ($\approx 0.5\%$),
which could be obscured by instrumental effects.
The principal challenge in accounting for systematics, however,
is deriving its exact effect on source count.
If this cannot be done, then one cannot write down a likelihood function.
However, Simulation-Based Inference (`SBI'),
otherwise known as likelihood-free inference,
uses a family of techniques to perform Bayesian statistical inference
when a likelihood function is intractable or unknown \citep{cranmer+20}.
This solves the problem of inverse inference where
generative models are extremely complex and have no obvious likelihood.\footnote{%
  For a list of recent studies in cosmology and astronomy that
  have used SBI, see \url{https://simulation-based-inference.org}.
}

In this work,
we deploy a neural posterior estimator and a neural likelihood estimator
to infer the posterior distribution of the cosmic dipole in CatWISE.
In particular,
we implicitly encode the \instrument{Widefield Infrared Survey Explorer}'s 
\citep[\instrument{`WISE'};][]{wright2010}
scanning law into our forward simulations,
demonstrating that this reproduces the reported ecliptic bias.
Since they are part of our generative process,
we do not need to `correct' or `reduce' the data before performing measurements.
We then explore a family of models,
using the learned likelihood from the \red{neural likelihood estimator}
to compute the Bayesian evidence for each model.

Our paper is structured as follows.
In Section~\ref{sec:background},
we give the background surrounding measurements of the cosmic dipole
and the current status of the dipole tension.
In Section~\ref{sec:catwise},
we discuss the CatWISE sample and how the original authors of S21
approached its analysis.
For Section~\ref{sec:method}, we describe our approach to the sample,
including how we produce forward simulations and construct our neural estimators.
We give our results in Section~\ref{sec:results},
which we discuss and conclude on in Section~\ref{sec:discussion}.

\section{Background}
\label{sec:background}

\citet{ellis1984} originally proposed that our motion through the Universe
induces a dipole in the counts of radio galaxies.
The key idea is that this movement brightens and concentrates 
galaxies ahead of us as long as we make a uniform cut in flux density.
Assume that we can describe the cumulative luminosity function of some galaxy survey
with a simple power law, so $ N(>S) \propto S^{-x} $.
Also assume that the spectral energy distribution of the galaxies can
be approximated with  another power law following their frequency of emission: 
$ S_{f} \propto f^{-\alpha } $.
Purely from the predictions of special relativity,
flux densities are transformed to our moving frame
according to $ S_{f}' = S_{f}\delta ^{1+\alpha} $
for $ \delta = \gamma (1 + \beta \cos \theta $ )
with Lorentz factor $ \gamma  $ and $ \beta = v_{\text{obs.}} / c $ 
(our speed in units of $ c $).
Meanwhile, relativistic aberration reduces the element of solid angle according to
$ d\Omega' = d\Omega \delta ^{-2} $.
The net effect, to first order, is a dipole in the distribution of radio galaxy counts
\emph{above some flux density}.
This dipole has an amplitude
\begin{equation}
  \mathcal{D} = (2 + x(1 + \alpha )) \beta ,
  \label{eq:eb_amplitude}
\end{equation}
which we refer to as the cosmic dipole amplitude
or the Ellis \& Baldwin (EB) amplitude.
The properties of the \red{galaxies at the flux limit, namely $ x $ and $ \alpha $,}%
\footnote{\red{See e.g. \citet{vonHausegger2024} for discussion on why these quantities
are defined near the limiting flux density in the \citet{ellis1984} test.}}
plus the use of $ v_{\text{obs.}} = v_{\text{CMB}} \approx 370 $ km s$^{-1} $
leads to an expected dipole amplitude $ \mathcal{D}_{\text{CMB}} $
given the kinematic hypothesis of the CMB dipole.
In principle,
we need not be restricted to radio surveys
and can perform this test across the EM spectrum.
\red{
  However, we require that the mean spectrum of the galaxies in the sample
  follows a power law within the passband used to define the flux limit, which
  is less obviously true for, e.g., quasars. This is because quasar spectral
  energy distributions exhibit thermal and non-thermal continua as well as
  strong emission lines, among other effects. Nonetheless, on aggregate across
  all redshifts, the mean spectrum is smooth \citep{secrest_coloq+25}.
}

Multiple studies using independent datasets have reported a cosmic dipole amplitude
that is inconsistent with the kinematic hypothesis.
The common element is that the reported amplitude is too large by at least a factor of 2.
If $ \mathcal{D}_{\text{measured}} > \mathcal{D}_{\text{CMB}}$,
this implies our motion is faster than the 370 km s$^{-1} $ derived from the CMB.
We defer the reader to \citet{secrest+25} for a more thorough review,
though a handful of recent results in the radio spectrum include:
\citet{wagenveld2023, oayda2024, wagenveld+25, bohme+25}.
These all show evidence for an excessive radio dipole amplitude
\citep[but see][]{wagenveld2024}.
In the near-infrared,
S21 and S22 reported a $ \approx 5 \sigma  $
discrepancy between the CatWISE dipole amplitude and the CMB amplitude,
which was confirmed in \citet[][`D23']{dam2023} and again in \citet{land-strykowski+25}.
This discrepancy with CatWISE is the chief focus of this work.
One intriguing piece of the picture is that, as presented in S22
and confirmed with a Bayesian approach in \citet{land-strykowski+25},
the results from the NRAO VLA Sky Survey \citep[NVSS;][]{nvss-survey} and CatWISE
are remarkably consistent with each other,
showing that this dipole excess persists across radically different wavelengths.
Meanwhile, in the optical regime,
issues owing to extinction near the Galactic plane and low source counts
mean that the question of an excessive dipole amplitude is not well-decided,
though it appears the dipole direction is aligned with the CMB dipole
\citep{mittal2024,mittal+24-erratum}.

The dipole tension has profound implications for our cosmological understanding,
since it would point against the assumption of homogeneity and isotropy.
Thus, verifying its authenticity is critical.
This motivates further inquiry into the effect of data systematics 
and the choice of statistical framework
\citep[for recent analyses of these issues, see e.g.][]{oayda+25,mittal+25}.
In this work,
we focus specifically on the observed ecliptic systematic in CatWISE,
unpacking why the density of quasars diminishes at higher \red{ecliptic latitudes}.
This ecliptic bias was discussed recently in \citet{abghari2024},
where the authors state that the `origin of the... gradient is unexplained'.
While S21 and S22 gave possible explanations that we explore in Section~\ref{sec:catwise},
simulations are essential in unravelling the potentially subtle effects at play.
This is especially so since 
it is not guaranteed that applying an ad hoc, a posteriori correction
--- that is, \red{weighting} cell counts according to a linear relation --- 
will remove the effect of the ecliptic systematic 
and the power it could have in the dipole mode.
Nor is it guaranteed that the CatWISE ecliptic bias can be exactly
parametrised by a quadrupole, as was assumed in \citet{panwar+24}.
However, before generating simulations,
we need sufficient understanding of the nuances of CatWISE.

\section{The CatWISE2020 Sample}
\label{sec:catwise}
The CatWISE sample in S21 was created from the CatWISE2020 data release \red{\citep{marocco2021}},
generated from observations using \instrument{WISE}.
\instrument{WISE} surveyed at wavelengths of
3.4~$\mu$m, 4.6~$\mu  $m, 12~$ \mu  $m, and 22~$ \mu  $m,
corresponding to the $ W1 $, $ W 2 $, $ W 3 $ and $ W 4 $ photometric bands respectively.
Before determining the dipole in CatWISE,
S21 made a number of cuts and selections on the raw sample
with the aim of extracting quasars and correcting for possible systematics.
First, the colour cut $W1 - W2 \geq 0.8$ is known to isolate objects
with AGN-dominated emission \citep{stern2012}.
Second, to account for reddening from Galactic dust in this mid-infrared sample,
the authors performed a manual correction to the $W1$ and $W2$ photometric magnitudes,
employing the Planck dust map \citep{planck2014} and extinction coefficients from \citet{wang2019}.
Third, the authors constructed a number of masks to mitigate poor-quality
photometry and image artifacts;
these masks were centred around resolved nebulae and bright stars,
as well as remaining spurious areas identified by the authors.
This, coupled with the Galactic plane mask for latitudes
between $-30^\circ$ and $30^\circ$, means that more than half the sky was masked.
The Galactic plane mask was justified on the basis
that number counts drop due to source confusion (S22).
Lastly, a bright magnitude cut of $ W 1 > 9 $ was used to mitigate saturation,
and while the $ W 1 < 16.4 $ cut introduces the cosmic dipole,
it was also chosen to safeguard against uneven source density 
due to \instrument{WISE}'s scanning law.

Despite these efforts,
CatWISE still exhibits a clear systematic shift in source density;
the count of quasars appears to diminish near the ecliptic poles
and increase along the ecliptic equator.
This induces a strong quadrupole ($\ell = 2$) signal in the density map,
exhibited in Fig.~\ref{fig:catwise_comparison}.
S22 gave two possible causes.
Firstly, deblending issues between fainter and brighter sources might be heightened
where the coverage is deeper (i.e., at the ecliptic poles),
leading to a drop in completeness.
Secondly, sources might be scattering into the colour cut
via Eddington bias \red{\citep{eddington+13}}.
More explicitly, where the photometric uncertainty is higher,
this scattering past the colour boundary is enhanced and leads
to an increase in source density.
Whatever the cause,
S21 corrected for this systematic shift in source density
by making a linear fit to the binned source counts as a function of declination,
then \red{using the inferred slope as a weighting function to remove the linear dependence.}
Thus, in S21 and S22,
this linear correction was \red{assumed to reflect the ecliptic systematic.}

This raises a number of questions.
First,
is it reasonable to assume that this systematic bias
— whatever its mechanism —
is fully (or accurately) described by a linear function?
Second, is there a way to probe the exact instrumental mechanism which
induces the ecliptic bias?
We propose that we can address these questions
by using a simulation-based statistical framework.
Indeed,
we contend that the most principled approach is to embed the systematic
as part of the data-generating process, leveraging the power of simulations
to learn its effect on the data.
One one hand,
we can heuristically deduce some functional dependence between
a nuisance parameter and the desired measurable
— like photometric uncertainty and source density.
Then, in a frequentist-style analysis,
the data might be scaled \red{or weighted} to correct for the effect (as in S21 and S22).
Alternatively, in a Bayesian approach,
this heuristic might be part of the model itself
(see e.g. the ecliptic bias parameter $\gamma_{\text{ecl}}$ in D23).
This allows inference when the explicit function relating the
nuisance parameter and the measurable is not known.
However, it forces a parametric fit
that might not reflect the physics at play.
Instead, if we know the mechanism which lies behind the systematic,
then we can write an explicit data-generating function that
maps model parameters to data — i.e., a simulation.
The power of SBI
is that the posterior distribution for the model parameters
can be learned without knowing the form of the likelihood function.
This grants access to all the tools of Bayesian statistics.

\section{Method}
\label{sec:method}
\subsection{Simulation function}
\begin{figure*}
  \centering
  \includegraphics[width=\columnwidth]{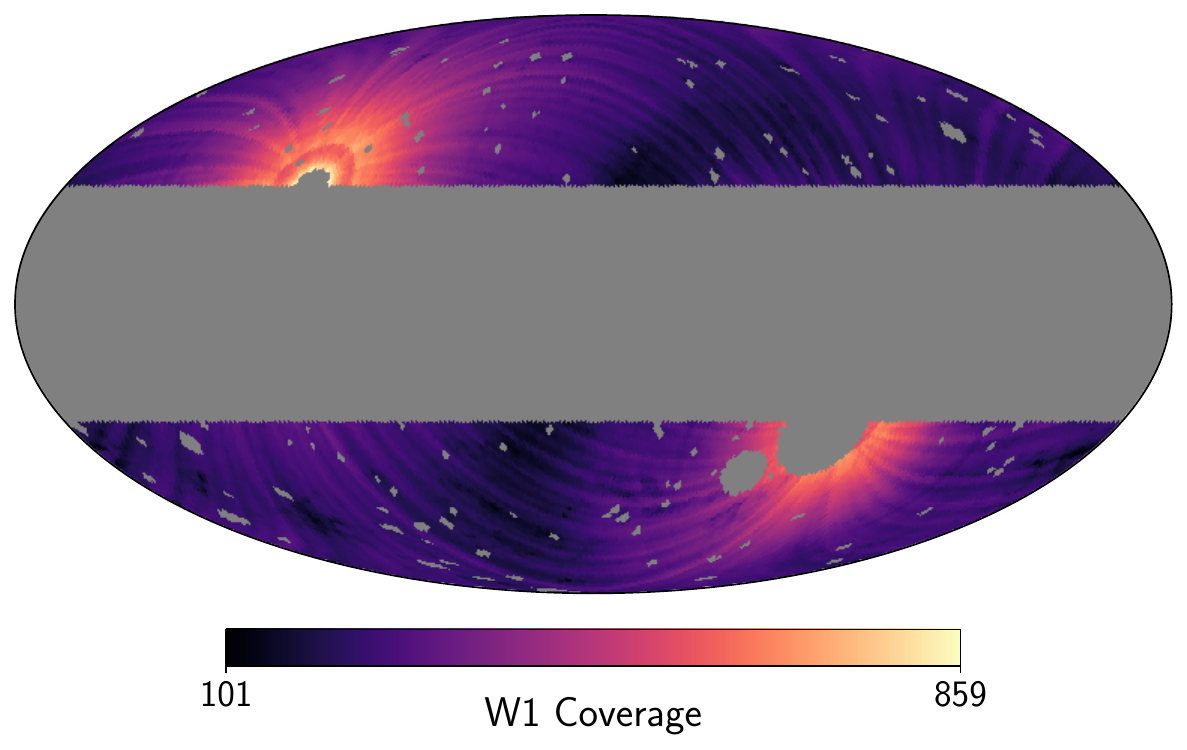}\hfill
  \includegraphics[width=\columnwidth]{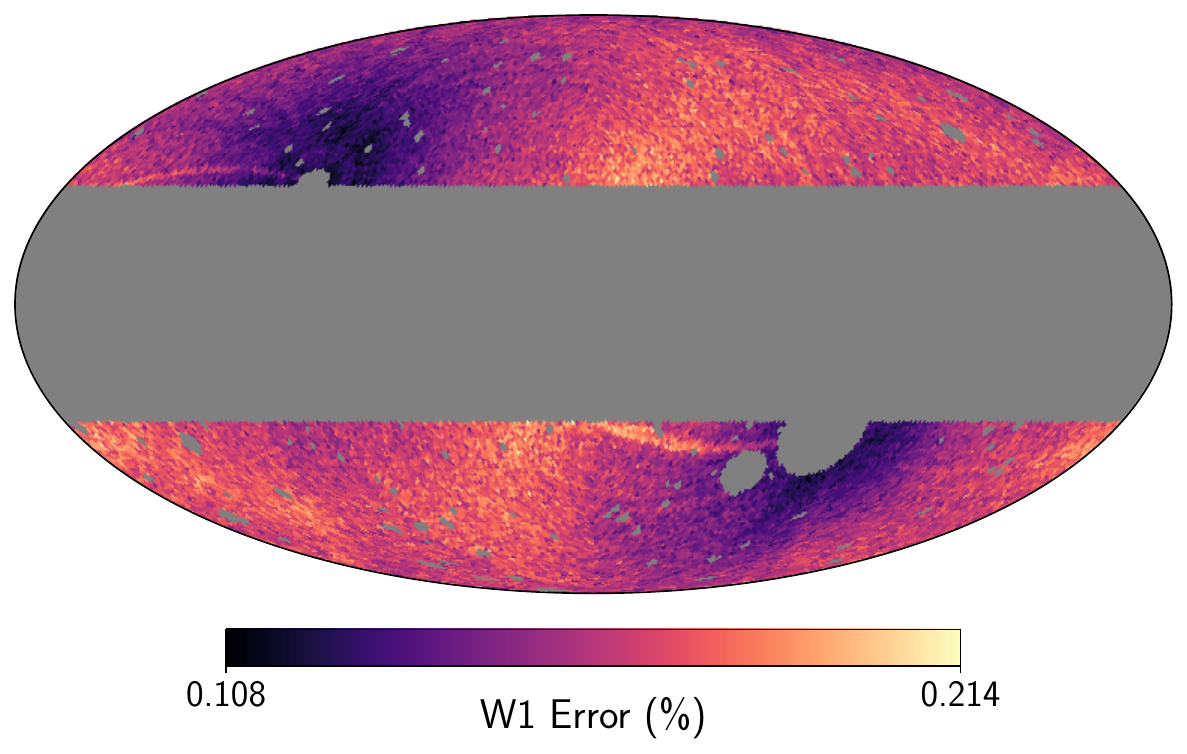}
  \caption{%
    Visual indication of the relationship between coverage and photometric error.
    A higher coverage implies a lower error.
    \textit{Left:} Median coverage per pixel for CatWISE in the $W1$ photometric band.
    \textit{Right:} Median photometric uncertainty (percentage error) per pixel 
      for CatWISE in the $W1$ band.
  }\label{fig:catwise_photo_error}
\end{figure*}

We first have to deduce a functional mapping from
the dipole parameters, $\Theta$, to the simulated CatWISE data, $\mathbf{D}$.
We assume that there are two essential features:
the cosmic dipole and the ecliptic systematic.
To create the latter feature,
we need knowledge about the physical or instrumental mechanism
inducing the change in source density.
As was pointed out in S22 and \citet{abghari2024},
the ecliptic trend appears to be correlated with \instrument{WISE}'s coverage
in the $W1$ and $W2$ bands.
Now, one expected effect of elevated coverage is a reduction in
photometric uncertainty. 
In patches of sky that \instrument{WISE} visits more often,
more photons are collected,
decreasing the per-pixel uncertainty with the inverse square root of the coverage
\citep{cutri+12}.
As evidence, Fig.~\ref{fig:catwise_photo_error} shows how the median $W1$
percentage error and the median W1 coverage change in different bins over the sky.
Compare this with the actual smoothed CatWISE source density map,
as shown in Fig.~\ref{fig:catwise_comparison}.

Given these visual cues, we postulate
that the ecliptic trend in source density can be explained
by sources scattering past the CatWISE cuts to a greater extent
where the photometric uncertainty is highest,
or where the coverage is lowest.
This is the Eddington bias as referenced in the previous section.
The essential ingredients are a steep luminosity function,
photometric magnitude measurements with some uncertainty and a cut in magnitude
\citep{teerikorpi+04}.
With these, it becomes more probable for faint sources 
outside the limit to scatter into the magnitude (flux density) cut
than for bright sources within the limit to scatter outside it.
Then, if we have an overall photometric uncertainty varying with sky position,
the magnitude of this effect would also vary with position.
Note that this bias would not only scatter sources into
the $W1 - W2 \geq 0.8$ colour cut,
but also into the $W1 < 16.4$ magnitude cut.
Accordingly, this complex effect would have magnitude, colour and spatial dependence,
making simulations essential for verifying if it can explain the ecliptic bias.

We give a sketch of the main elements of our forward simulation here,
which we explain in further detail in the following sections:
\begin{enumerate}
    \item Generate $N$ 2D samples from the joint empirical distribution
        for $W1$ and $W2$. These are the true source magnitudes.
    \item For each sample,
        compute the $W1 - W2$ colour and lookup up the corresponding spectral index $\alpha$.
    \item For each sample, choose a position uniformly on the sphere.
    \item For some heliocentric speed $v_{\text{obs.}}$
      and direction ($l^\circ$, $b^\circ$), apply special relativity        
      by Doppler boosting the $W1$ and $W2$ magnitudes and aberrating the source positions.
    \item Discard sources which fall within the \red{masked region of the sky}.
    \item For each boosted magnitude, add a photometric uncertainty depending
        on the source magnitude itself and the coverage at the associated point in the sky.
    \item Make the cuts $9 < W1 < 16.4$ and $W1 - W2 > 0.8$ on the boosted magnitudes.
    \item Count the number of sources in equal-area pixels on the sky.
\end{enumerate}
While this order roughly follows our simulation routine,
we do not follow that order in the sections below.
Also, from here on, we refer to our simulated CatWISE sample as `CatSIM'
while the term `CatWISE' refers to the empirical sample from S21.

\subsubsection{Drawing photometric samples}
\label{sub:drawing_photometric_samples}
We start by constructing a less conservative CatWISE2020 sample,
imposing the cuts $W1 < 17.0$ and $W1 - W2 > 0.5$.
We refer to this as the `deeper' CatSIM.
Because our ansatz is that sources outside the magnitude-colour cut
scatter into it due to photometric uncertainty,
our deeper sample needs to accommodate this possibility.
To obtain this sample,
we proceed as in S21,
querying the NASA/IPAC Infrared Science Archive for CatWISE2020 sources with
photometric uncertainty in $W1$ and $W2$ greater than 0,
$W12$ colour greater than 0.5 and $W1$ magnitude less than 17.
We then apply the exact same dust and astrometric corrections
as in S21.\footnote{
  We rely on the script `correct\_catwise.py' from the paper's code
  at \url{https://zenodo.org/records/8303800}.
}
From this parent sample,
we create an empirical $W1$-$W2$ 2D histogram and use the inferred probabilities
per bin to generate  new samples from the distribution.
The number of samples we draw, $ N_{\text{init.}} $,
is simply a parameter we can fit.
Now, in principle this parent distribution already contains the dipole signal;
that is, magnitudes have been boosted and de-boosted due to our motion.
We assume that this effect is averaged out over the forward and reverse hemispheres
such that we can use the distribution to draw rest frame magnitudes.
\red{
  Further, while we anticipate that the distribution will be somewhat broadened
  by photometric uncertainty, since the percentage errors are typically of
  $ \mathcal{O}(1\%) $ or less, the effect will be negligible.
}

\begin{figure}
  \begin{center}
    \includegraphics[width=\columnwidth]{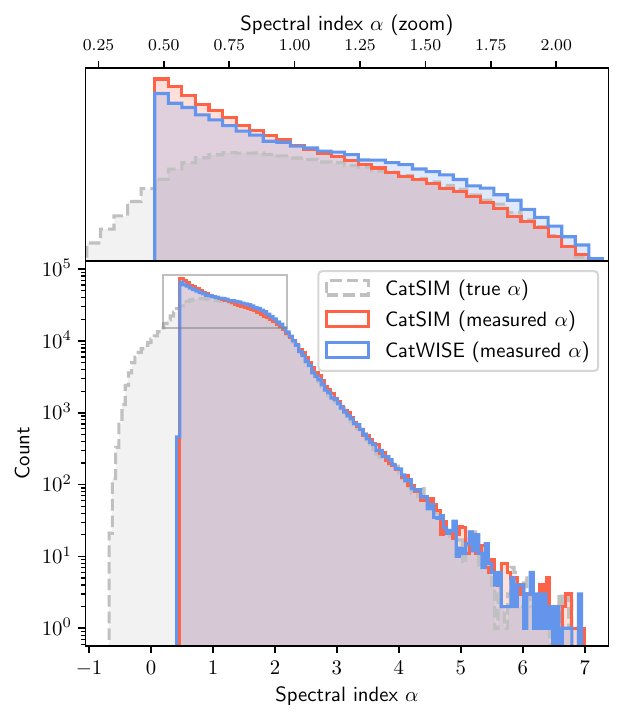}
  \end{center}
  \caption{
    The spectral indices determined from our simulation (CatSIM)
    and the empirical data (CatWISE). The dashed grey histogram indicates the true indices
    in the simulated sample, while the solid red histogram indicates what an observer
    would measure after noise has been added to the photometric magnitudes.
    The blue histogram indicates the distribution of $\alpha$ from the actual CatWISE dataset.
    In the top panel, we zoom in on the peak of all three histograms to highlight
    the difference between the simulated and empirical indices.
  }\label{fig:catwise_alpha}
\end{figure}

Next, for each photometric sample, we compute the $W1 - W2$ colour and
look up the corresponding spectral index $\alpha$.\footnote{%
  We use the `alpha\_colors.fits' file from the code for S21
  to infer the relation between colour and $\alpha$.
}
This is based on the assumption that the spectral energy distribution of each
source follows a power law, so $S_f \propto f^{-\alpha}$.
We refer to these spectral indices as the `true' spectral indices.
This is because while the spectral index is an intrinsic property of the source
explicitly determined from its spectral energy distribution,
the actual measured spectral index is a function of the measured colour,
for which there is an uncertainty.
Indeed, since we know that the photometric uncertainty varies over the sky 
(see Fig.~\ref{fig:catwise_photo_error}),
so too will the uncertainty in colour and thereby the uncertainty in spectral index.
We can glean the effects of this from Fig.~\ref{fig:catwise_alpha}.
There, we plot the `true' spectral indices from CatSIM, 
as indicated by the dashed grey histogram.
These are the spectral indices of each simulated source determined \emph{before}
adding any error to the magnitudes.
We also plot the `measured' spectral indices in CatSIM,
as indicated by the solid red distribution.
These are the spectral indices determined 
\emph{after} adding position-dependent uncertainty to the colour,
as a real observer would see (see Section~\ref{sub:adding_uncertainty}).
Note that while both distributions are consistent for high $\alpha$,
below $\alpha \approx 2$ they diverge,
with the true spectral indices being dispersed past $\alpha \approx 0.5$.
The hard cut for the measured distribution corresponds to the colour cut
that is applied after introducing photometric error (see Section~\ref{sub:cuts}).
In essence, the uncertainty in colour means sources with true spectral indices
below the cut `leech into' the sample; thus, the measured indices are
inconsistent with the true indices.
This accords with the Eddington bias ansatz --- more sources below
the colour cut (lower $\alpha$) are being scattered to higher colours 
(higher $\alpha$) above the cut.

We also compare the CatSIM spectral indices to the empirical distribution
of $\alpha$ from the real CatWISE sample, 
as indicated by the solid blue histogram in Fig.~\ref{fig:catwise_alpha}.
This is essentially identical to figure 2 in S21
except for there being $\mathcal{O}(10^4)$ less sources here due to our slightly modified mask
(see Section~\ref{sub:cuts}).
While the measured CatSIM indices are broadly consistent with the real indices,
there appears to be a slight divergence for $0.5 < \alpha < 2$,
which we zoom into at the top panel.
The distribution of CatWISE indices has a more significant `twist'
than the simulated indices,
\red{%
  meaning that the simulated indices (red) are more numerous than the CatWISE
  indices (blue) for $ \alpha < 1 $, whereas this is flipped for $ 1 < \alpha < 2 $.
}
Although this does suggest our simulation is not fully representing the original
sample, the effect is likely to be small.
The mean for the real indices is $\bar{\alpha} \approx 1.26$,
whereas the mean for the simulated measured indices is $\bar{\alpha} \approx 1.23$.
Although our simulation does not require the EB amplitude \eqref{eq:eb_amplitude}
to be computed, for the sake of comparison,
this difference in $\alpha$ amounts to 
$\Delta \mathcal{D} = 6.5 \times 10^{-5}$ at $x=1.75$ and $ v=v_{\text{CMB}} $,
or roughly $0.9\%$ of the anticipated dipole amplitude.

\subsubsection{Sampling sky positions}
For each W1-W2-$\alpha$ sample generated by the above procedure,
we choose a point uniformly over the surface of a sphere.
This implicitly assumes that there is no dependence of either of these three variables
on sky position \red{(as expected under the cosmological principle)},
apart from the ecliptic bias and cosmic dipole we introduce later.
This gives each source a position $(l^\circ, b^\circ)$ in Galactic coordinates.

\subsubsection{Special relativity}
Now, we add the effect of our motion.
This introduces three dipole parameters:
the observer's speed $ v_{\text{obs.}} $,
the dipole direction in Galactic longitude $ l^\circ $
and the dipole direction in Galactic latitude $ b^\circ $.
For each source position,
we apply relativistic aberration where the angle is transformed as
\begin{equation}
  \cos \theta' = \frac{\beta + \cos \theta}{\beta\cos\theta + 1}.
  \label{eq:aberration}
\end{equation}
Here, $\theta'$ is the angle between the direction of motion (the dipole vector)
and the source in the observer's moving frame, $\theta$ is that same angle
in the source rest frame and $\beta = v_{\text{obs.}} / c$.
This introduces the $\delta^2$ factor from the \citet{ellis1984} equation
for the integral source counts.
We then boost the rest-frame magnitudes $m_f$ directly;
since $S'_f = S_f \delta^{1 + \alpha}$, we can write
\begin{equation}
  m'_f = m_f - 2.5 (1 + \alpha) \log_{10} \delta.
  \label{eq:boost_magnitude}
\end{equation}
We boost both the $W1$ and $W2$ magnitudes using \eqref{eq:boost_magnitude}.
Altogether, these two steps imprint the signal of our motion into CatSIM.

\subsubsection{Adding uncertainty}
\label{sub:adding_uncertainty}
Next, we associate each source with a photometric uncertainty $\sigma_{WX}$.
To do this, we assume that a source's uncertainty
is some function of \red{its passband magnitude $m_{WX}$ and its passband coverage
$C_{WX}$.}
\red{
  We compute the coverage in either band by creating median $ C_{\text{WX}} $
  sky maps using the empirical CatWISE dataset 
  (as in the left pane of Fig.~\ref{fig:catwise_photo_error}),
  ultimately allowing us to convert a source's position to a coverage given
  WISE's scanning law.
  Now, in Figure~\ref{fig:cov_mag_error_heatmap},
  we visualise the magnitude-coverage relationship for the W1 passband.
  The median photometric error in each cell follows a reasonably smooth
  relationship between magnitude and the logarithm of the coverage,
  apart from the region in the top right of the upper panel.
  We explain this difference in Section~\ref{sub:cuts}.
  While in Figure~\ref{fig:cov_mag_error_heatmap} we show the relationship in W1
  only, it is important to note that the W1 and W2 photometric uncertainties
  are highly correlated.
  It is insufficient to, independently in each passband,
  compute a photometric error given a source's magnitude and coverage; this
  should also depend on the error in the other passband.
  This is especially relevant since we make a cut in W12 colour:
  a selection that we anticipate is the progenitor of CatWISE's ecliptic bias.
  Accordingly, we construct a four-dimensional lookup using the empirical
  CatWISE data.
  We bin sources according to their magnitude and coverage in both passbands,
  then in each bin record the joint distribution of W1 and W2 photometric errors.
  Thus, given a source's simulated boosted magnitude $ m'_{\text{WX}} $ and its
  logarithmic coverage $ \log _{10} C_{\text{WX}} $, where $ X \in \{1, 2\} $,
  we can determine a photometric uncertainty $ \sigma _{\text{WX}} $.
}

With $\sigma_{WX}$, we draw an error for each source 
assuming Gaussian errors in magnitude space:
\begin{equation}
  \Delta m'_{WX} \sim \mathcal{G}(\mu = m'_{WX}, \sigma = \sigma_{WX}).
  \label{eq:mag_error}
\end{equation}
Thus the error added onto each source
is not deterministic and will vary
at each function call.
We will return to the assumptions in \eqref{eq:mag_error} later
when we consider models with different noise properties.

\subsubsection{Cuts, binning \& masking}
\label{sub:cuts}
\begin{figure}
  \begin{center}
    \includegraphics[width=0.9\columnwidth]{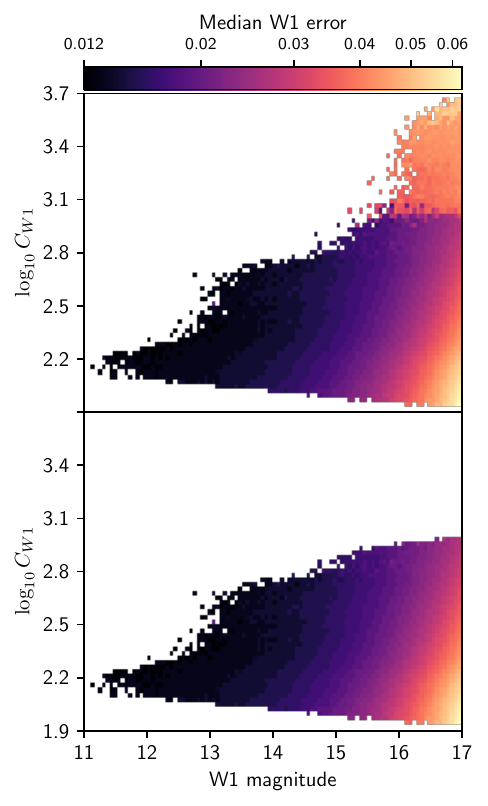}
  \end{center}
  \caption{
    Relationship between coverage, magnitude and median error in the W1 band.
    In each coverage-magnitude bin,
    we compute the median photometric error across all real CatWISE samples
    made from the deeper catalogue ($W1 < 17.0$ and $W1 - W2 > 0.5$).
      \textit{Top:} without the northern ecliptic pole mask.
      \textit{Bottom:} with the northern ecliptic pole mask.
    }
    \label{fig:cov_mag_error_heatmap}
\end{figure}

Now that each source has a boosted position and photometric magnitude,
we impose the same cuts on the simulated sample as was used in S21.
In particular,
we select only sources with $9 < W1 < 16.4$ and $W1 - W2 > 0.8$.
We then bin sources into equal-area pixels
using the \software{healpix} algorithm,\footnote{\url{https://healpix.sourceforge.io/}}
as  implemented in the \textsc{python} package \textsc{healpy} \citep{Gorski2005, Zonca2019},
yielding a source density map. 
Lastly, we mask out pixels using the original mask defined in S21
and our own minor additions.
Specifically, we mask out an additional 5$^\circ$ around the north ecliptic pole.
We noticed this region has highly elevated coverage
and introduces a discontinuity in the otherwise smooth relationship between 
photometric error, coverage and magnitude we reference in Section~\ref{sub:adding_uncertainty}.
We show this in Fig.~\ref{fig:cov_mag_error_heatmap},
where the orange region at the top of the upper pane is removed
after masking the northern ecliptic pole (lower pane).
\red{%
  Our simulation cannot capture the typical photometric uncertainties in that
  region, which suggests something extrinsic to the relationship between
  magnitude and coverage affects the error there.
  That being said, we verified that our results are substantively unaffected by
  including the north ecliptic pole.
  We comment on this again in Section~\ref{sec:results}.
}

\subsubsection{Comparison to CatWISE}
\begin{figure*}
  \centering
  \includegraphics[width=\columnwidth]{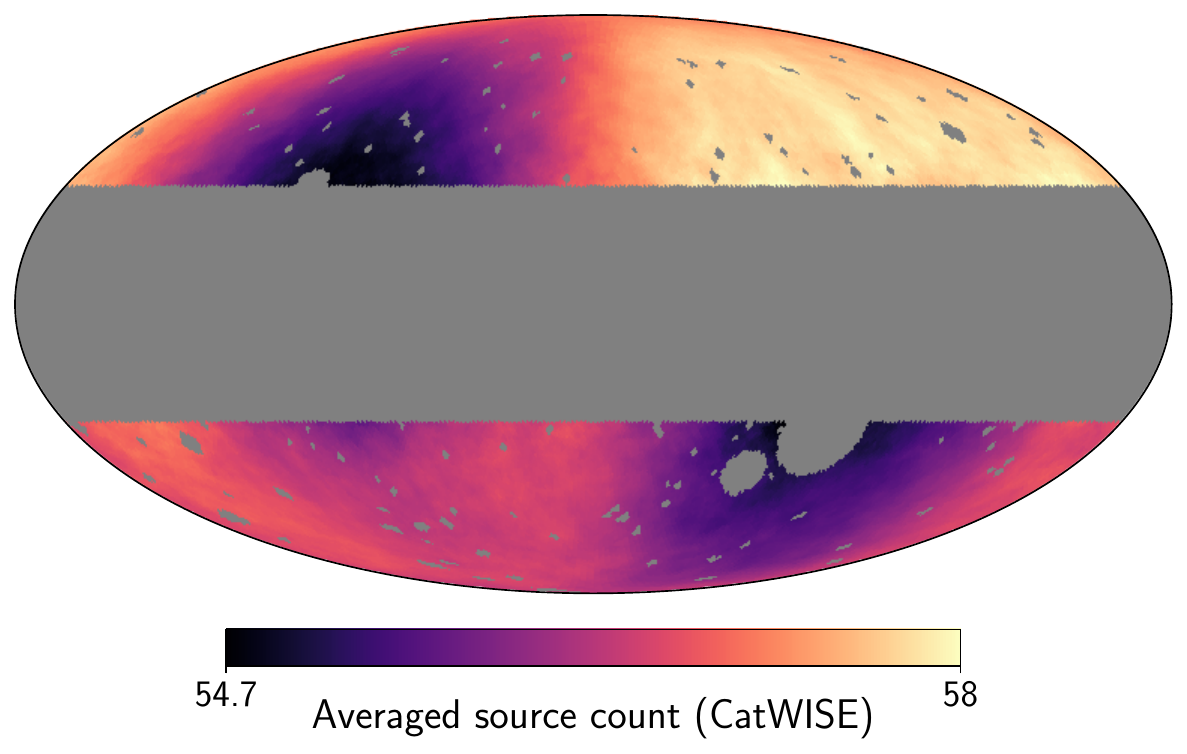}\hfill
  \includegraphics[width=\columnwidth]{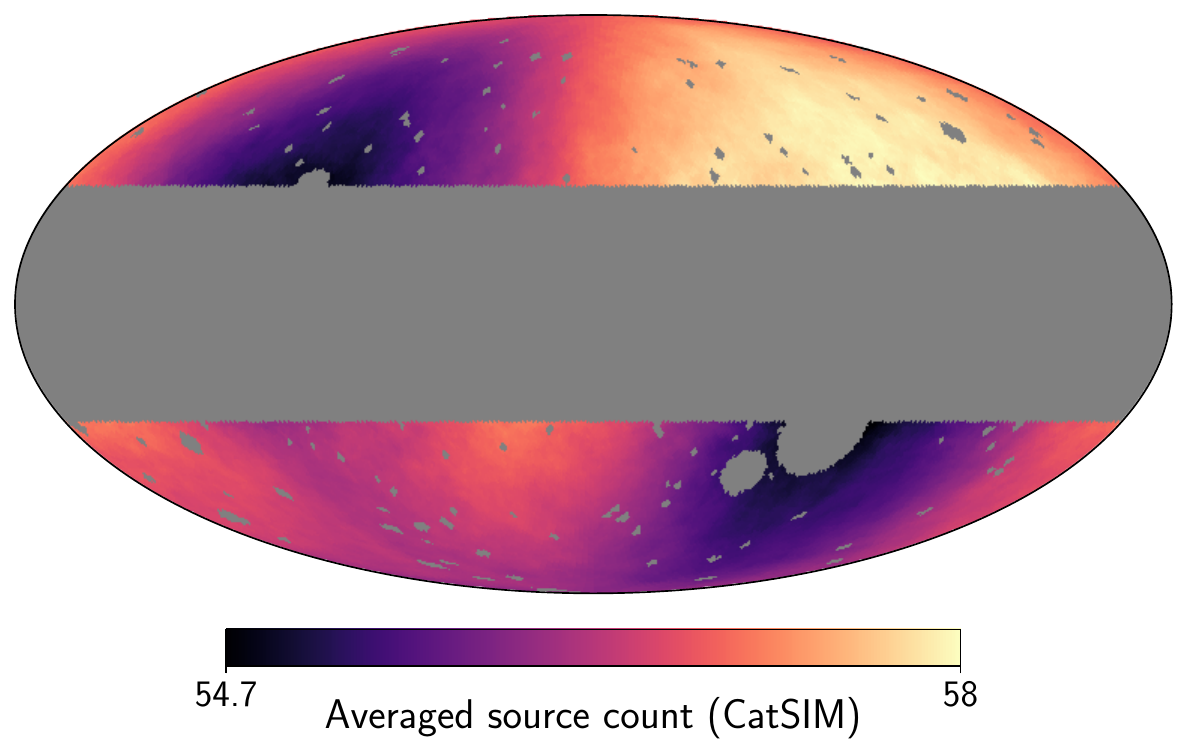}
  \caption{
    Comparing the real CatWISE data to the outputs of our simulation.
    \textit{Left:} The real CatWISE2020 sample (smoothed with a moving average)
    as similar to that in \citet{secrest2021} except for minor modifications to the mask.
    \textit{Right:} Example of a CatSIM sample (smoothed) generated from one function call.
      The particular arguments passed to the function are the median parameters
      from Fig.~\ref{fig:free_gauss_err_corner},
      \red{which includes an observer velocity of $ v_{\text{obs.}} \approx 740\,$km\,s$^{-1}$}.
      In this sense, the plot is also a posterior predictive check for that model.
  }\label{fig:catwise_comparison}
\end{figure*}

We compare the outputs of our simulation function (a CatSIM sample)
to the real CatWISE sample in Fig.~\ref{fig:catwise_comparison}.
The output density maps in both cases have been smoothed with a moving average
over a scale of 1 steradian.
To reiterate, each call of the function is not deterministic
since various steps of the function rely on random processes.
Nonetheless,
the similarity between the actual and simulated data is striking.
This illustrates that the ecliptic bias can be recreated
by simulating how photometric uncertainty varies with \textit{WISE}'s coverage,
which we take as verification of our Eddington bias ansatz.
There are some caveats to this, however,
in particular relating to our treatment of the width of CatWISE's
photometric error distribution.
This is elaborated on in Section~\ref{sec:discussion}.

The exact parameters chosen for the CatSIM function call in 
Fig.~\ref{fig:catwise_comparison}
are taken from the posterior distribution of one of the models we fit later,
specifically the model shown in Fig.~\ref{fig:free_gauss_err_corner}.
We stress that the model parameters \emph{do not} include the dipole amplitude.
As indicated in the steps outlined above,
all we require as an input is an initial number of samples to draw from
the W1-W2 histogram, an observer's speed as a multiple of the CMB-dipole-derived velocity,
and the direction of the observer's motion.
Thus, whether or not we compute the mean spectral index,
or find the correct slope of the luminosity function,
or `correct' for some kind of bias is irrelevant.
All of the physical behaviour at the heart of the \citet{ellis1984} effect
is implicitly encoded in our simulation.

\subsection{Simulation-Based Inference}
Having developed a data-generating function that accounts for \textit{WISE}'s scanning law,
we still need to learn the relationship between the model parameters and the data.
Traditionally,
we would write down a likelihood function and apply Bayes's theorem,
where the desired posterior distribution is
\begin{equation}
  P(\mathbf{\Theta} | \mathbf{D}, M) = \frac{\pi (\mathbf{\Theta} | M) \mathcal{L}(\mathbf{D} | \mathbf{\Theta}, M)}{\mathcal{Z}(\mathbf{D} | M)}.
  \label{eq:bayes}
\end{equation}
Here, $ \mathbf{D} $ is the data, $ \mathbf{\Theta} $ is the set of model parameters
and $ M $ is the model.
Meanwhile, $ \mathcal{L} $, $ \pi  $ and $ \mathcal{Z} $ refer to the
likelihood function, prior function and Bayesian evidence (marginal likelihood) respectively.
We need the Bayesian evidence for model comparison,
where, assuming models $A$ and $ B $ have a prior odds ratio of 1,
the Bayes factor is
\begin{equation}
  B_{AB}
    = \frac{P( M_{A} | \mathbf{D})}{P(M_B | \mathbf{D})}
    = \frac{\mathcal{Z}(\mathbf{D} | M_A)}{\mathcal{Z}(\mathbf{D} | M_B)}.
  \label{eq:bayes_factor}
\end{equation}
However, CatSIM does not elicit an obvious choice of likelihood function.
In previous works \citep[see e.g.][]{mittal2024,oayda2024},
we assume the cell counts follow a Poisson distribution,
the rate parameter of which varies over the sky according to a dipole.
While the cell counts for CatSIM are anticipated to be Poisson deviates,%
\footnote{%
  \red{See however \citet{vonhausegger2025},
  in which CatWISE's cell counts are shown to follow
  an over-dispersed or `general' Poisson distribution.}
}
we do not a priori know what the rate parameter at each cell will be.
Again, there are complex effects arising from \instrument{WISE}'s scanning law
that do not have an obvious mathematical relation.

\subsubsection{Learning likelihoods with SBI}
In the absence of a likelihood function,
we can leverage the power of neural networks as a density function approximator.
In essence,
given a set of realisations of CatSIM ($ \mathbf{D} $)
for certain model parameters ($ \mathbf{\Theta} $),
we want to learn either the likelihood function or the posterior distribution directly.
Normalising flows can achieve this \citep{papamakarios+19}. 
The essential idea is to transform a simple base distribution
to a more expressive probability distribution via a series of diffeomorphisms
(associated by parameters $ \phi $) that are learnt by the neural network.
The neural network's loss function is coded to the probability of the target variable
conditioned on another variable,
the exact target and conditioner depending on the architecture (see below).
If we denote the learnt distribution $ q_{\phi}$
(sometimes referred to as the surrogate posterior or surrogate likelihood),
then the network's output $ q_{\phi }(\mathbf{D} | \mathbf{\Theta}, M) $
or $ q_{\phi } (\mathbf{\Theta} | \mathbf{D}, M) $
converges to the true likelihood and posterior respectively
with increasing simulation count \citep{cranmer+20}.

In this work, we focus on two main approaches to neural density estimation:
the \red{neural posterior estimator} (`NPE') and the \red{neural likelihood estimator} (`NLE').
The NPE, originally proposed in \citet{papamakarios+16},
learns the posterior distribution: the probability of the model parameters
as conditioned on the data.
While an NPE allows one to efficiently generate posterior samples by calling
the neural network,
since we only have the learnt posterior $ P(\mathbf{\Theta} | \mathbf{D}, M) $
and our original prior $ \pi (\mathbf{\Theta} | M) $, by \eqref{eq:bayes} we only have access
to $ \mathcal{L} $ up to a normalising constant.
Importantly, this means the evidence $ \mathcal{Z} $ cannot be computed,
rendering Bayesian model comparison impossible.
On the other hand,
NLEs \citep{papamakarios+19-nle} directly learn the likelihood function.
While this adds the additional requirement of sampling the posterior distribution
with an algorithm like MCMC or nested sampling,
it means that both the evidence and posterior distribution are accessible,
opening up the full suite of Bayesian inference.

The label NLE or NPE is typically associated with \emph{amortized} approaches.
The idea is to generate a number of parameter samples from the prior likelihood function,
use these samples in the simulator function to generate many datasets,
then use the $ (\mathbf{\Theta}, \mathbf{D}) $ pair to learn the target distribution.
This has the advantage of generalising across many parameter realisations
from the full prior space,
such that the learned target distribution can be used to do inference
on many actual observations $ D_{i} $.
In this sense, the upfront cost of training the inferer is \emph{amortized over time}
as many different observations are quickly analysed \citep{zammit-mangion+25}.
However, this is suboptimal if one is only interested in a single target observation $ D_{0} $;
computational effort has been spread across the prior space,
but we only care about the single, small region of parameter space 
associated with $D_0$.
\emph{Sequential} methods (`SNPE'/`SNLE') solve this issue by iteratively updating the proposal
(prior) distribution over $ L $ rounds of inference \citep{papamakarios+19-nle}.
In the first round, we use $ \pi (\mathbf{\Theta}| M) $ to draw parameter samples
and generate simulations.
Once we have learnt the posterior distribution $ q_{\phi}^{(1)} $,
this becomes the proposal distribution for the next round of inference,
and so on.
After the final round, the distribution $ q_{\phi }^{(L)}  $ is the final result
we use for probabilistic inference.
Thus, at the same simulation count,
this sequential approach generally has better performance than the amortized
approach for one observation $ D_{0} $.
However, it cannot generalise as well across multiple observations.
Since we are only interested in the CatWISE sample as in S21
(see Fig.~\ref{fig:catwise_comparison}),
we use the sequential algorithm.

\subsubsection{Estimating the Bayesian evidence}
We note that while posterior and likelihood estimation with normalising flows
has seen extensive use in recent years,
direct estimation of the Bayesian evidence is rarer.
\citet{spurio-mancini+23} recently performed an extensive profiling of the
effectiveness of neural density estimators,
coupled with the learned harmonic mean estimator \citep{mcewan+21},
at computing $ \mathcal{Z} $.
\citet{bastide+25} similarly compared a number of algorithms to determine
$ \mathcal{Z} $, including harmonic-mean- and importance-sampling-based approaches.
In this work, we instead directly use the
final learned likelihood $ q_{\phi}^{(L)} \approx \mathcal{L} $ from our SNLE
coupled with a nested sampling algorithm to compute the final posterior and evidence.
Since our implementation is built in the JAX ecosystem \citep{jax2018github},
we rely on the Handley Lab's fork of the \textsc{blackjax} \textsc{python} library
\citep{cabezas2024blackjax, yallup2025nested},
which implements GPU-native nested slice sampling.\footnote{%
  \url{https://github.com/handley-lab/blackjax};\\
  \url{https://handley-lab.co.uk/nested-sampling-book/intro.html}
}
Now, as our SNLE yields both the posterior and evidence,
our SNPE is somewhat redundant.
However, it is effective as a consistency check for our results
as its architecture is quite distinct to that of the SNLE
(see Section~\ref{sec:appendix}).

In fact, we found in testing that the results of the SNLE and SNPE are extremely
sensitive to their architecture.
We highlight the salient challenges here and leave a more detailed exploration
for Section~\ref{sec:appendix}.
A critical concern is the dimensionality of the data.
The original CatWISE sample in S21 was created with $ N_{\text{side}} = 64 $,
amounting to $ 49\,152 $ healpixels.
If one naively trains an SNLE on this data,
they must accurately model a $ 49\,152$-dimensional base distribution:
a virtually intractable problem.
This is why crafting a summary statistic is usually essential
when applying an SNLE or SNPE to actual data.
However, using a summary statistic would break our use case.
To explain, suppose we apply some non-invertible, dimensionality-reducing
transform $ f_{\text{sum.}} $ such that $ \mathbf{z} = f_{\text{sum.}}(\mathbf{D}) $.
Then, we have
\begin{equation}
  \mathcal{Z}(\mathbf{D} | M) \neq \mathcal{Z}(\mathbf{z} | M).
\end{equation}
That is, the evidence associated with the compressed data $ \mathbf{z} $
is not generally equal to the evidence of the original data,
and there is no Jacobian we can write for the change of variables.
This would prevent model comparison using the explicit probability of the data
given the model.

We can circumvent this by taking advantage of the hierarchical nature of HEALPix data
using the `nested' ordering.
What we want to show is that, by performing some downscaling operation
on the data, we do not impact the inferred posterior distribution
or the relative marginal likelihoods for different models.
Now, if we start at $ N_{\text{side}} = 64 $,
then a pixel at the next coarsest resolution ($ N_{\text{side}} = 32 $) consists 
of a block of four interior pixels from $ N_{\text{side}} = 64 $.
Since our data describes counts of objects inside a pixel,
the value of the cell at the coarser resolution is just the sum of the counts
of the four pixels at the finer resolution.
If we assume the counts themselves are Poisson deviates,
then the distribution of the coarse pixel counts will also be Poissonian.
We only need to take care of masked pixels:
they are excluded from the sum, and a coarse pixel containing four masked
sub-pixels is itself masked.
The net result is that we can take a map of arbitrary $ N_{\text{side}}$
and downscale to a desired resolution while keeping an explicit likelihood function.

Concretely, denote the `resolution' of the \software{healpix} map as $ \ell $
such that the number of pixels at some resolution is $ N_{\ell} = 12 (4^{\ell} ) $,
meaning $ N_{\text{side}} = 2^\ell $.
We call the resolution of the initial map $ \ell_\text{high} $ 
and the target low resolution $ \ell_{\text{low}} $.
Also, we index the low resolution pixels with $ i_{\ell_\text{low}} $,
the pixels with the next highest resolution (i.e., twice the $ N_{\text{side}} $)
as $ i_{\ell_\text{low}+1} $, and so on.
Thus, we refer to the observed count of a pixel at some resolution $ \ell $ as $ k_{i_{\ell}} $.
We denote the set of all child pixels or sub-pixels for some pixel $ i_{\ell} $
at resolution $ \ell $ as 
$ C_{i_{\ell}} = \{ i_{\ell+1} \}_{\text{child, }i_{\ell}} $,
since they necessarily have resolution $ \ell+1 $.
To represent the masking procedure described in the previous paragraph,
we introduce a binary mask $ m_{i_\ell} \in \{ 0, 1\} $
where 1 means the pixel is unmasked and 0 means it is masked.
With this, we can define a downscaling operation
$ D_{\ell} : (\mathbf{k}_{\ell}, \mathbf{m}_{\ell}) \mapsto (\mathbf{k}_{\ell-1}, \mathbf{m}_{\ell-1})$,
mapping the vector of pixel counts $ \mathbf{k}_\ell $ and binary mask $ \mathbf{m}_\ell $
to the next coarsest resolution:
\begin{equation}
  k_{i_{\ell - 1}} = \sum_{j \in C_{i_{\ell - 1}}} m_j k_j \\
\end{equation}
and
\begin{equation}
  m_{i_{\ell-1}} =
  \begin{cases}
  1, & \text{if } \exists\, j \in C_{i_{\ell-1}} \text{ such that } m_j = 1, \\
  0, & \text{otherwise.}
  \end{cases}
\end{equation}
This means a full downscaling operation from $ \ell_\text{high} $ to $ \ell_{\text{low}} $
is just the composition of these individual downscales:
\begin{align}
  (\mathbf{k}_{\ell_{\text{low}}}, \mathbf{m}_{\ell_{\text{low}}}) 
    &=  
      (D_{\ell_{\text{low}+1}} \circ \cdots \circ D_{\ell_{\text{high}}})
      (\mathbf{k}_{\ell_{\text{high}}}, \mathbf{m}_{\ell_{\text{high}}}) \\
    &\coloneq D_{\ell_{\text{high}} \to \ell_{\text{low}}} 
      (\mathbf{k}_{\ell_{\text{high}}}, \mathbf{m}_{\ell_{\text{high}}}).
  \label{eq:downscale_operation}
\end{align}
Then, the log likelihood is the sum of the per-pixel
Poisson probabilities given an anticipated rate parameter $ \lambda $:
\begin{equation}
  \ln \mathcal{L}_{\ell_{\text{low}}} = \sum _{i_{\ell_\text{low}}}^{N_{\ell_{\text{low}}}}
    \text{Pois} \left( 
      k_{i_{\ell_\text{low}}}
      \Big|
      \, \lambda_{i_{\ell_\text{low}}}
    \right)
    \label{eq:downscale_lnlike}
\end{equation}
where $ k_{i_{\ell_{\text{low}}}} $ and $ \lambda_{i_{\ell_{\text{low}}}} $
are obtained using the downscale operator in \eqref{eq:downscale_operation}.

Again, our SNLE \emph{learns} a likelihood, 
so we have no need to (and cannot) write one down.
However,
the above exercise shows that we can downscale mock CatSIM samples 
and assess the accuracy of our SNLE
by benchmarking with the explicit likelihood in \eqref{eq:downscale_lnlike}.
Thus, we can be assured that downscaling the data
\emph{has no substantial impact on the results}.
While the exact value of the evidence will be different,
we can verify that the posterior distributions and Bayes factors
are identical whether or not we use the high or low dimensional map.
We show that this is indeed the case in Section~\ref{sec:snle_accuracy}.

In the SNLE arm of our SBI pipeline,
after generating CatSIMs $ \mathbf{D} $,
we apply this downscaling process to produce reduced datasets $ \mathbf{z} $.
The neural network sees these downscaled maps, not the original CatSIM maps.
For the SNPE,
the foregoing concerns are not relevant because the target variable
is no longer the data but the model parameters $ \mathbf{\Theta} $.
As long as we do not use non-invertible transform on the parameters,
the learned posterior distribution can be conditioned on summarised data.
Thus, we adopt a CNN-style approach,
performing 1D convolutions on the healsphere as in \citet{krachmalnicoff+19}.
The CNN's weights and biases are updated alongside that of the normalising flow,
effectively allowing the network to learn a summary representation of the data
while evaluating the posterior distribution.
We fully unpack our choices of hyperparameters in Section~\ref{sec:appendix}.

With our SNLE, we have all the tools needed for standard Bayesian inference.
Our learned likelihood $q_\phi^{(L)}$ is the likelihood in Bayes's theorem \eqref{eq:bayes},
so we use it to compute the posterior distribution and the Bayesian evidence
for a suite of models.
Meanwhile, our SNPE produces --- via very distinct architecture --- another 
posterior distribution, enabling verification of our SBI pipeline.

\subsection{Models tested}
\label{sub:models_tested}
We computed the Bayesian evidence for the following models,
the exact characteristics of which we explain further below.
\begin{itemize}
  \item Free dipole, extra error, Gaussian
  \item Free dipole extra error, Student's $t$
  \item Free dipole, no extra error, Gaussian
  \item Free dipole, no extra error, Student's $t$
  \item CMB direction, free velocity, extra error, Gaussian
  \item CMB velocity, free direction, extra error, Gaussian 
  \item CMB velocity \& direction, extra error, Gaussian
  \item Dipole from S21, extra error, Gaussian
  \item Dipole from D23, extra error, Gaussian
\end{itemize}

\subsubsection{Free dipole}
The label `free dipole' means a model in which the dipole parameters
($ \hat{v}_{\text{obs.}}$, $l^\circ$, $b^\circ $)
are drawn from the following prior functions instead of being constrained to a particular value:
\begin{align}
  \hat{v}_{\text{obs.}} &\sim \mathcal{U}[0, 8] \\
  l^\circ &\sim \mathcal{U}[0, 360] \label{eq:prior_l} \\
  b^\circ &\sim \mathcal{P}[-90, 90] \label{eq:prior_b}.
\end{align}
We introduce $ \hat{v}_{\text{obs.}} = v_{\text{obs.}} / v_{\text{CMB}}$, i.e. 
the observer's speed in units of the CMB-derived velocity.
$ \mathcal{U}[a,b] $ denotes a continuous uniform distribution between $ a $ and $ b $,
so, for example, we impose a prior of an observer speed between 0 and 8 times
$ v_{\text{CMB}} $.
$ \mathcal{P} $ denotes a `polar distribution' where $ P(b) = \frac{1}{2} \sin b$
for Galactic latitude in radians $ [0, \pi ] $.

We also need a number of initial sources to populate the celestial sphere with,
which we denote as $ N_{\text{init.}} $.
This is somewhat analogous to the monopole parameter $ \overline{N} $ usually used
when fitting for the cosmic dipole:
a high $ N_{\text{init.}} $ means more sources end up in the final density map.
However, most sources initially drawn will be discarded by the final cuts.
We adopt the prior $ \log _{10} N_{\text{init.}} \sim \mathcal{U}[7.477, 7.602] $,
corresponding to a log uniform prior between 30 and 40 million.

\subsubsection{Error distribution}
The label `extra error' and `no extra error' refers to how we handle the
photometric uncertainties generated for CatSIM.
As mentioned, we lookup a photometric error given a source's simulated magnitude and coverage
(see Fig.~\ref{fig:cov_mag_error_heatmap} and Section~\ref{sub:adding_uncertainty}).
These are derived from the deeper variant of the real CatWISE dataset,
before the final colour and magnitude cuts.
In testing, we found that the simulated decrease in source density at the ecliptic poles
was too small as compared with what we actually observe in CatWISE.
We discuss the implications of this later, but for now,
we can choose to add an extra error term in quadrature
which enhances the photometric error:
\begin{equation}
  \sigma_{WX\text{, final}}^2 = \sigma_{WX}^2 + \eta_{\text{extra}} \, \sigma _{WX}^2.
  \label{eq:extra_error}
\end{equation}
The $ \eta_{\text{extra}} $ term simply increases the width of the distribution we sample errors from.
If $ \eta_{\text{extra}} = 0 $, we have `no extra error', and the final error 
\red{purely comes from the formal uncertainties} quoted in the
empirical CatWISE data (see Fig~\ref{fig:cov_mag_error_heatmap}).
Now, we assume $ \eta_{\text{extra}}  $ is the same for both the W1 and W2 magnitudes;
that is, $ \eta _{W 1,\text{ extra}} = \eta _{W 2,\text{ extra}} $.
We originally used a unique $ \eta_{\text{extra}}  $ for each band,
though we later uncovered substantial degeneracy between the two parameters.
For this parameter, we adopt a prior $ \eta_{\text{extra}} \sim \mathcal{U}[0, 8] $.

We also vary the assumed distribution for these errors.
Models with the label `Gaussian' assume normally-distributed errors,
whereas those with the label `Student's $t$' assume errors from that distribution.
This means the Student's $t$ models need an extra term for the shape parameter
of the distribution, which we denote as $ \xi $.
At high $ \xi $, the distribution is approximately Gaussian;
at low $ \xi $, the tails of the distribution are dramatically enhanced,
allowing more sources with `anomalously' large errors.
\red{%
  We adopt a prior $ \log _{10} \xi \sim \mathcal{U}[0.3, 3] $,
  where the lower limit is predicated on the fact that the variance of the
  Student's $ t $ distribution is infinite for $ 1 < \xi \leq 2 $ and undefined
  for $ \xi \leq 1 $.
}

\subsubsection{CMB models}
For the CMB models,
we fix the dipole parameters to their CMB-derived values.
For example, `CMB velocity, free direction' means $ v_{\text{obs.}} = v_{\text{CMB}} $,
or $ \hat{v}_{\text{obs.}} = 1 $,
while the direction is sampled from \eqref{eq:prior_l} and \eqref{eq:prior_b}.
Therefore, the `CMB velocity \& direction' model is just a dipole completely
consistent with the CMB dipole.
Note that we also assume Gaussian errors for the CMB models with the
extra error term $ \eta_{\text{extra}} $, as described above.
For future reference, 
we drop the `extra error' and `Gaussian' descriptors for these models.

\subsubsection{Literature models}
We also fix our dipole parameters to those inferred from S21
and D23, where for the latter study we use the median parameters
from the posterior distribution (shown by figure 2 therein).
Like the CMB models, we also assume Gaussian errors with the $ \eta $ term,
and we drop the `extra error' and `Gaussian' descriptors.

\section{Results}
\label{sec:results}
\begin{figure*}
  \begin{center}
    \includegraphics[width=0.95\textwidth]{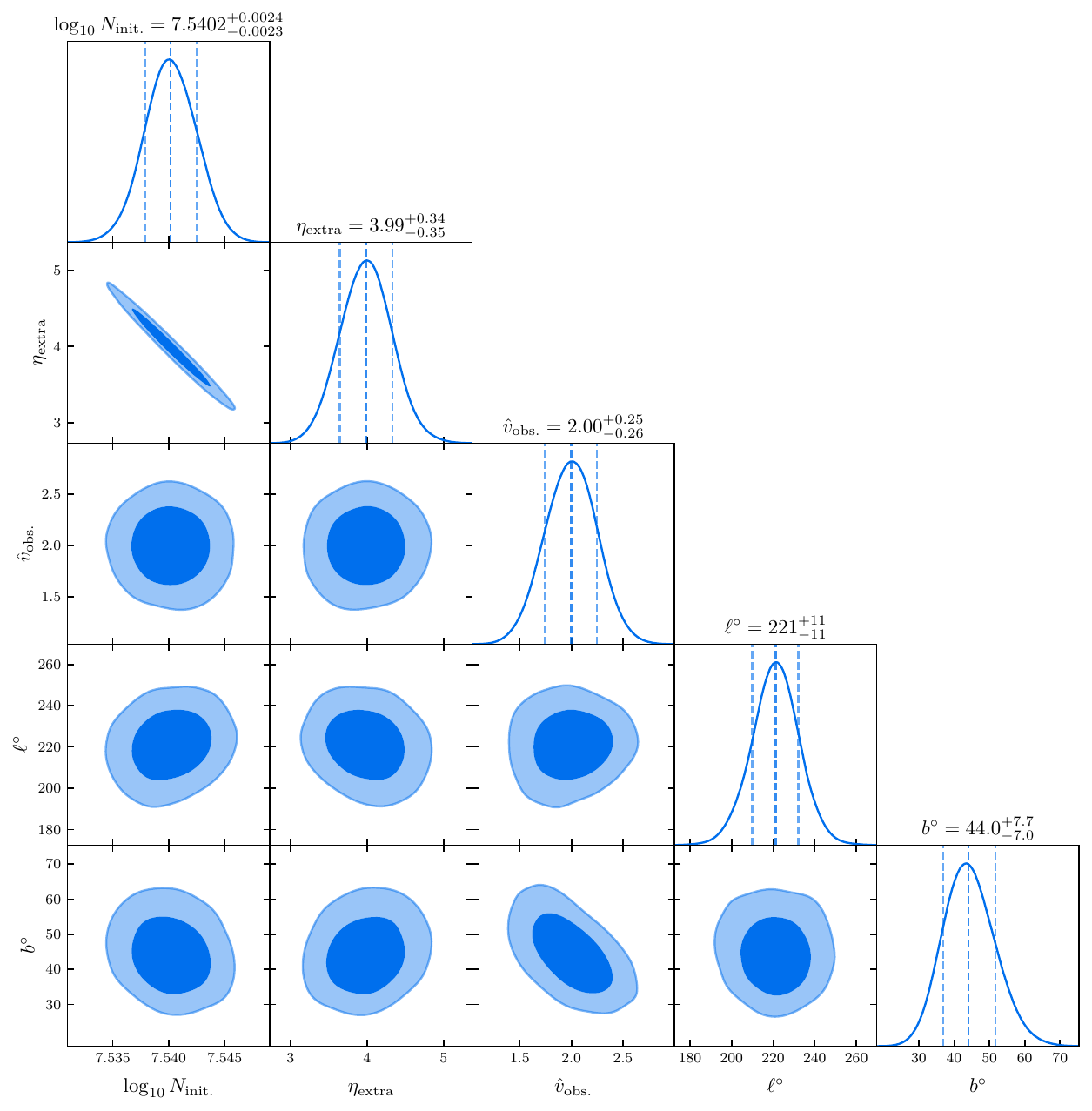}
  \end{center}
  \caption{
    The inferred posterior distribution for our fiducial model:
    `free dipole, extra error, Gaussian'.
    This assumes a dipole with free parameters 
    and an $ \eta_{\text{extra}} $ term which increases the width of CatWISE's
    photometric error distribution (assumed to be Gaussian).
    A $ 1 \sigma  $ credible interval is indicated by the dashed lines
    and titles of the 1D marginals.
    Meanwhile, $ 1 \sigma  $ and $ 2 \sigma  $ intervals are shown by the filled contours
    in the 2D marginals.
  }\label{fig:free_gauss_err_corner}
\end{figure*}

\begin{figure}
  \begin{center}
    \includegraphics[width=\columnwidth]{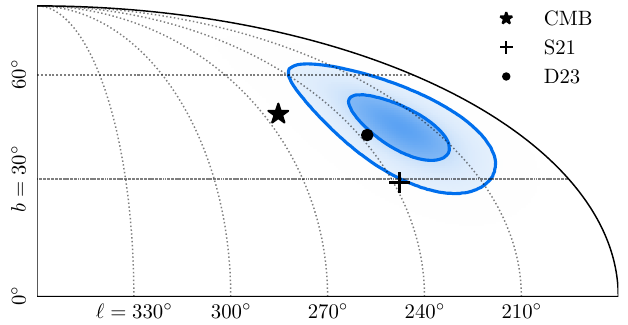}
  \end{center}
  \caption{
    Inferred dipole direction for the `free dipole, extra error, Gaussian' model,
    shown in Galactic coordinates.
    The contours enclose $ 1 \sigma  $ and $ 2 \sigma  $ intervals of posterior density
    for a 2D Gaussian.
    The direction of the CMB dipole is shown by the star,
    while the results of \citet[][`S21']{secrest2021} and \citet[][`D23']{dam2023}
    are indicated by the plus and dot respectively.
    }\label{fig:free_gauss_extra_err_sky}
\end{figure}

We show the log Bayes factors for each model in Table~\ref{tab:bayes_evidence_table}.
\begin{table}
    \centering
    \caption{
      Summary of the log Bayes factor of each model with respect to 
      our fiducial model (`free dipole, extra error, Gaussian').
      Thus by definition the log Bayes factor of that model is 0.
      As mentioned in Section~\ref{sub:models_tested},
      the \citet{secrest2021} and \citet{dam2023} models,
      as well as the CMB-based models,
      assume the `extra error' parameter defined in \eqref{eq:extra_error}.
  }
    \label{tab:bayes_evidence_table}
    \red{\input{logb_summary.tex}}
\end{table}
These are written with respect to the `free dipole, extra error, Gaussian' model,
which we set as our fiducial model.
Thus, by definition, this has a log Bayes factor of 0.
We also show the corner plot for this model in Fig.~\ref{fig:free_gauss_err_corner}.
We then project the $ l^\circ $-$ b^\circ $ 2D marginal distribution onto the
sky in Fig.~\ref{fig:free_gauss_extra_err_sky}, illustrating the
direction of the inferred dipole.
This posterior distribution was generated using our NLE,
and we show it is consistent with the outputs of the NPE in 
Fig.~\ref{fig:corner_npe_vs_nle} of Section~\ref{sec:npe_verification}.
\red{
  We also verified that we reproduce the results in Fig.~\ref{fig:free_gauss_err_corner}
  if we choose to leave the north ecliptic pole unmasked
  (see Section~\ref{sub:cuts} for discussion on this).
  In particular, we find
  $ \hat{v}_{\text{obs.}} = 2.05\substack{+0.24\\-0.25} $ 
  and $ l^\circ, b^\circ = 221\substack{+11\\-11}, 40.8\substack{+7.1\\-6.4} $.
}

To generate the uncertainties for the log Bayes factors in Table~\ref{tab:bayes_evidence_table},
we note that after roughly three rounds of inference,
the posterior distribution computed from the NLE is the same for subsequent rounds
while the estimate of $ \ln \mathcal{Z} $ fluctuates.
This is because we instantiate the neural network from scratch at each round,
thereby randomly re-initialising the weights and biases.
This introduces another level of stochasticity to the sensitive evidence estimate.
We compute one standard deviation on these $ \ln \mathcal{Z} $ values.

\red{
  Given the uncertainties in $ \ln B $, we note that
  the `free dipole, extra error, Gaussian' model has similar support to
  the `free dipole, extra error, Student's $ t $' model.
  However, the common parameters inferred in either case are virtually identical
  (we deduce $ \hat{v}_{\text{obs.}} = 2.12\substack{+0.24\\-0.24} $ 
  and $ l^\circ, b^\circ = 218\substack{+11\\-11}, 41.4\substack{+6.8\\-5.9} $
  for the Student's $ t $ variant; cf. Fig.~\ref{fig:free_gauss_err_corner}).
  We also observed degeneracy between the Student's $ t $ shape parameter $ \xi $
  and the extra error term $ \eta _{\text{extra}} $, which could illustrate that this model
  is not the most parsimonious.
  Thus, we are justified in selecting the `free dipole, extra error, Gaussian'
  model as our fiducial model for comparison with other studies.
}

\section{Discussion \& Conclusions}
\label{sec:discussion}
A dipole consistent with the CMB dipole is overwhelmingly disfavoured
\red{($ \ln B = -9.9 \pm 1.2 $).}
This means there is substantially stronger preference for our fiducial dipole
with parameters as shown in Fig.~\ref{fig:free_gauss_err_corner}.
This dipole has an amplitude that is twice as large as the CMB expectation
\red{($ \hat{v}_{\text{obs.}} = 2.00\substack{+0.25\\-0.26} $
or roughly $ 740\,$km\,s$^{-1} $ )},
and is thus consistent with the factor of two excess in S21.
The direction is $ \approx 3 \sigma $ away from the CMB direction 
(Fig.~\ref{fig:free_gauss_extra_err_sky}),
a higher level of disagreement than the just under $ 2  \sigma  $ tension
reported in S21 (see figure 4 therein).
Altogether,
this is substantial evidence that the dipole in CatWISE remains discrepant
with the CMB dipole --- even after reproducing the ecliptic bias in our simulations.

The model with the globally highest Bayesian evidence is the S21
dipole, which is mildly preferred over our fiducial model \red{($ \ln B = 3.1 \pm 1.2 $).}
The next highest evidence is yielded from a model assuming the dipole from D23
\red{($ \ln B = 2.4 \pm 1.0 $)},
although the large uncertainties on $ \ln B $ means this is, at best, marginally preferred.
We interpret this as signifying that CatSIM is still consistent
with the studies that used the ecliptic bias correction
(although again note that these models use the extra error term $ \eta _{\text{extra}} $).
It is worth mentioning, though, that our inferred amplitude is
slightly lower than S21 (they recovered $ \hat{v}_{\text{obs.}} = 2.16 $),
and reasonably lower than D23 ($ \hat{v}_{\text{obs.}} = 2.68 \pm 0.23 $).
As revealed in Fig.~\ref{fig:free_gauss_extra_err_sky},
while our recovered direction is consistent with D23,
it is mildly discrepant with S21.
That being said,
these minor differences do not tip the model odds in favour of our fiducial model.

Speaking generally, Table~\ref{tab:bayes_evidence_table} reveals that models
which use the extra error term defined in \eqref{eq:extra_error}
are most favoured.
Indeed,
the `free dipole, no extra error, Gaussian' model has
catastrophically less evidence than our fiducial model;
\red{$ \ln B = -110.5 $} implies a model odds ratio of \red{$ \approx 1.0 \times 10^{-50} $}!
This means that the ecliptic bias cannot be fully replicated unless one
assigns additional error to the CatWISE photometric magnitudes.
Indeed, the value of \red{$ \eta _{\text{extra}} = 3.99\substack{+0.34\\-0.35} $}
from Fig.~\ref{fig:free_gauss_err_corner} means that the errors have to be 
\red{slightly more than} doubled.
There are two interpretations of this result:
either the uncertainties in the published CatWISE sample are genuinely underestimated,
or there is an additional physical process --- not present in our simulation ---
that produces the exact same effect as Eddington bias.
S22 speculated that heightened sensitivity to faint sources where the coverage is higher
could also introduce deblending issues with brighter sources inside the cut.
This would effectively drop the source counts.
Now, this very well could be part of the picture,
though it is not immediately obvious it would affect the counts in the same
way as Eddington bias.
For one, Eddington bias scatters sources into the cut more often where coverage is lower.
Meanwhile, a deblending bias would instead reduce the probability of source
detection in bins with higher coverage.
Perhaps this does induce the same drop in source density at the ecliptic poles,
though this cannot be said asserted until it is encoded into CatSIM.

Speaking of CatSIM,
our principal task was to investigate whether Eddington bias
is responsible for the ecliptic trend in CatWISE.
We have given strong evidence that it is highly relevant.
However, our CatSIM could go further. 
In addition to the deblending issue mentioned above,
one might also want to apply the astrometric and dust corrections
as part of the simulation itself.
We mention these in Section~\ref{sub:drawing_photometric_samples},
and in our case, we simply corrected the deeper sample identically to S21
to enable a more faithful comparison with that study.
However, a more principled way to frame this is not as a correction
we must make before measuring the data,
but an implicit feature which is accounted for at the level
of probabilistic inference.
In other words, it is encoded into the data-generating process.
While beyond the scope of this work,
future studies might seek to investigate the degree to which dust contributes to
a change in source density, especially considering the issues in
\citet[][though the sample was in the optical regime,
\red{in which extinction is expected to have a much larger effect than in the infrared}%
]{mittal2024,mittal+24-erratum}.
\red{To our knowledge, no study has as of yet considered whether the dust correction
employed in S21's CatWISE sample affects the inferred dipole.}

As this paper was submitted,
\citet{vonhausegger2025} released a new re-analysis of CatWISE.
There, the authors parametrically modelled the dipole and higher order multipoles
in the quasar sample, deploying a full Bayesian framework.
In their models, the ecliptic bias was accounted for either with the
linear ecliptic factor $ \gamma _{\text{ecl}} $, as from \citet{dam2023},
or with a quadrupole.
The authors also postulated that Eddington bias
is responsible for the ecliptic trend,
and showed that since the inferred value for $ \gamma _{\text{ecl}} $
drops with brighter magnitude cuts,
it is consistent with the systematic being related to the fractional photometric error,
or the sample's typical signal to noise ratio.
Our work gives direct evidence for this proposition.

Although our result is similar to S21, S22
and D23,
we emphasise that understanding the data --- and the instrumental
systematics that complicate it --- is paramount.
The rise of massively-parallel computation means that simulations are cheap.
Moreover, machine learning with SBI allows one to be confident that these systematics
and their potentially unforeseeable consequences 
are not neglected when making statistical inferences. 
This is critical for the next era of cosmological datasets,
which will shed light on the cracks that may be appearing in $ \Lambda $CDM
\citep{di-valentino}.

In summary, our finding of an anomalous dipole in CatWISE,
taken with the radio galaxy studies,
represents increasingly clear evidence that something is amiss with our
understanding of these samples,
or (more seriously) that our interpretation of the CMB dipole is wrong.
To be confident of this last proposition,
future studies of the cosmic dipole will need to carefully consider
the instrumental effects that are at play.
The tools we have developed and applied for the first time to the cosmic dipole problem
are easily extendible to other galaxy surveys.
However, we will need to inquire deeper into the systematic effects
that can impact the rate of source detection.
We stress that this is not optional but a necessity.

\section*{Acknowledgements}
\red{
  We thank the anonymous referee for their insightful comments which improved
  this paper's quality. This research has made use of the NASA/IPAC Infrared
  Science Archive, which is funded by the National Aeronautics and Space
  Administration and operated by the California Institute of Technology.
}
The corner plots in this work were generated with the \software{getdist} library
\citep{lewis+25}.
As mentioned in the text,
this study made extensive use of the \software{surjectors} library \citep{dirmeier+24},
as well as the Handley Lab's fork of \software{blackjax},
available at \url{https://github.com/handley-lab/blackjax}.
We thank the contributors of these software libraries.
We also extend our gratitude towards: 
the authors of \citet{secrest2021} for making their code publicly available;
Brendon Brewer for helpful discussions regarding modelling 
the photometric error distributions;
and Will Handley and Harry Bevins for insightful discussions
on SBI and nested sampling,
as well as for hosting OTO and GFL at the Kavli Institute in Cambridge.
OTO is supported by the University of Sydney Postgraduate Award.

\section*{Data Availability}

The data used in this study will be made available with
a reasonable request to the authors.



\bibliographystyle{mnras}
\bibliography{refs} 




\appendix

\section{SNLE/SNPE architecture}
\label{sec:appendix}
After extensive testing,
we settled on the \software{surjectors} library \citep{dirmeier+24}
for the implementation of our NLE and NPE.
This library uniquely implements surjective layers,
breaking the typical invertibility requirement for normalising flow layers
by `discarding' the data to a decoder distribution that is learned during gradient descent.
Overall, we found that performance was improved using a surjective layer
even after downscaling the CatSIM maps to $ N_{\text{side}} = 4 $.

For each arm of our SBI pipeline, we use a total of $ 50\,000 $ simulations
spread over 15 sequential rounds of inference.
We adopt a 90\%/10\% training-validation split with a batch size of 100.
We also use an AdamW optimiser with a learning rate of 0.0001 for the NLE
and 0.001 for the NPE.

\subsection{NLE}
Starting from the raw CatSIMS $ \mathbf{D} $ and their associated parameters $ \mathbf{\Theta} $,
we first $ z $-score the data across the batch axis,
yielding a per-healpixel mean and standard deviation.
For the model parameters, we perform a global $ z $-score, except for the dipole
direction which we transform to Cartesian $ \hat{x} $, $ \hat{y} $ and $ \hat{z} $.
Again, since the model parameters are the conditioning variable in the NLE arm,
there is no need to keep track of Jacobian terms.

Next, we use a series of masked autoregressive flow layers (MAFs)
as defined in \citet{papamakarios+17}.
The bijector function is simply a scalar affine transform,
and our conditioner function is a masked autoencoder \citep[MADE;][]{germain+15}
with 4 layers, each with 256 neurons and a tanh activation function.
We use a total of four MAFs, after each of which we place a permutation 
reversing the data order, before a surjective later.
Specifically, the surjection is an \verb|AffineMaskedAutoregressiveInferenceFunnel|
as implemented in \software{surjectors}.
This uses the same masking procedure as in the MAFs,
but drops 50\% of the data to a latent space using an affine transform
and a conditioner functions with 4 layers of 128 neurons.
We use a Gaussian for the decoder distribution.
After considering the masked pixels,
this drops out 56 data dimensions of our 112D data vector,
leaving a 56D-vector that propagates through a final chain of 6 MAFs.

\subsection{NPE}
Unlike our NLE,
the NPE learns a distribution with a significantly lower dimensionality.
This is just the number of parameters for each model,
far less than the input $ N_{\text{side}} = 4 $ map for the NLE.
Accordingly, we use a smaller architecture
and deploy 5 MAFs (2 layer / 64 neuron conditioner).
We also use the `atomic loss' or NPE-C algorithm as proposed in \citet{greenberg+19}.
In terms of the data,
we again $ z $-score but do this globally, computing a single
mean and standard deviation across all pixels.
For the parameters, we use a prior bijector,
mapping from constrained prior space to unconstrained latent space 
(which the normalising flow sees) via a sigmoid function.

Further, we use a CNN embedding network as part of the training.
After the data is normalised,
we perform a series of 1D convolutions, 
reducing the high resolution CatSIM map to a lower dimensional representation.
Because of memory constraints, we first downscale the $ N_{\text{side}} = 64 $ map
to $ N_{\text{side}} = 32 $.
Then, we apply the convolutions as described in \citet{krachmalnicoff+19},
which we port to our \software{jax} implementation.
This essentially uses a kernel of length 9 with a stride of 9
on an unravelled \textsc{healpy} map such that the convolution is performed
on all the immediate neighbours of a healpixel (see figure 3 therein).
After this convolution, we average pool in a nested block of four pixels,
halving the $ N_{\text{side}} $.
Thus, we perform a total of three convolutional steps to move from
$ N_{\text{side}} = 32 \to N_{\text{side}} = 4 $.
In order, we use a 2, 4 and 8 filters for these convolutions.
After this, we pass the reduced data through
a multi-layer perceptron with two layers at 128 neurons each
and a dropout rate of 0.2,
which then maps to an output vector of 32 dimensions.
This output vector is what ultimately conditions the NPE;
since the embedding net is included in the inference pipeline,
its weights and biases are updated as the posterior distribution is estimated.

\section{NLE accuracy verification}
\label{sec:snle_accuracy}
To ensure our NLE produces accurate estimates of the true Bayesian evidence,
we of course need to benchmark against a known likelihood.
We thus take \eqref{eq:downscale_lnlike} and generate downscaled pseudo-CatSIM
maps according to the downscaling procedure described there.
A pseudo-CatSIM map has an identical mask to CatWISE/CatSIM and a similar source count,
however is not simulated with Eddington bias.
In some sense this is a `raw dipole' map,
and is generated by drawing a Poisson deviate for cell $ i $ assuming a rate
parameter $ \lambda_i = \overline{N}(1 + \mathcal{D} \cos \theta_i) $,
where $ \overline{N} $ is the mean density, $ \mathcal{D} $ is the EB amplitude
and $ \theta _{i} $ is the angle between the dipole vector and cell.
We start from an $N_{\text{side}} = 64$ sample with a dipole described by an
observer speed of $\hat{v}_{\text{obs.}} = 2$ (units of CMB speed),
corresponding to an EB dipole amplitude of $\mathcal{D} = 0.01$,
and a direction of $(l^\circ, b^\circ) = (215, 40)$.
We then downscale the map to $N_{\text{side}} = 4$.

\begin{figure}
  \begin{center}
    \includegraphics[width=\columnwidth]{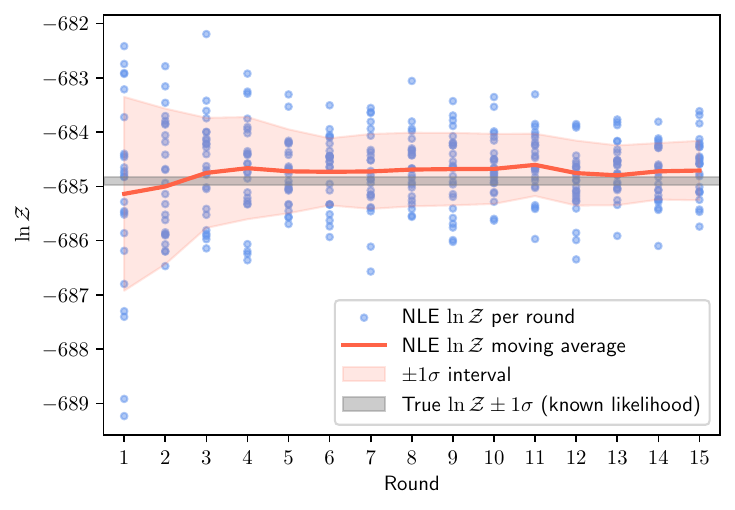}
  \end{center}
  \caption{
    Learned NLE evidence as a function of inference round for a pseudo-CatSIM map
    with a known true $ \ln \mathcal{Z} $ (grey horizontal strip).
    The evidence is computed for 25 independent runs 
    (estimates shown by the blue points).
    A moving average with a window of 2 is shown by the red line,
    which is enclosed within a $ 1 \sigma  $ envelope.
  }\label{fig:evidence_evolution}
\end{figure}

We first compute the true Bayesian evidence of the downscaled map via \eqref{eq:downscale_lnlike}.
We then compare this to the NLE-estimated evidence as a function of NLE round.
The results are given in Fig.~\ref{fig:evidence_evolution},
where we show the scatter in estimated evidences from 25 independent runs of the NLE.
At later rounds in the inference process,
the NLE-estimated evidence appears to converge to the true evidence,
with the true evidence sitting well within the $1\sigma$ deviation
on either side of the moving average.
Also, to illustrate the accuracy of the inferred posterior distribution,
in Fig.~\ref{fig:corner_evidence_acc} we overlay the true posterior 
computed using \eqref{eq:downscale_lnlike}
on top of that inferred from the NLE for one of the runs.
There is excellent agreement between the two.

\begin{figure}
  \begin{center}
    \includegraphics[width=\columnwidth]{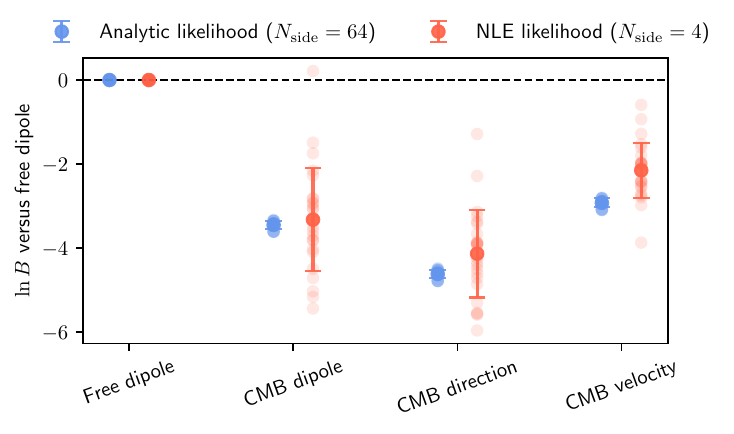}
  \end{center}
  \caption{
    Bayes factors inferred by our NLE (red; right) versus known true Bayes factors
    (blue; left).
    The individual estimates across 25 independent NLE runs are shown by the
    transparent red points, the mean and standard deviation of which
    are given by the red error bar.
    For the NLE estimates, we also apply downscaling to $ N_{\text{side}} = 4 $
    --- thus the actual value of $ \ln \mathcal{Z} $ is different for the
    NLE versus the analytic reference (the data is different).
    Nonetheless, we illustrate that the Bayes factors are consistent.
  }\label{fig:lnB_scatter}
\end{figure}

\begin{figure*}
  \begin{center}
    \includegraphics[width=0.9\textwidth]{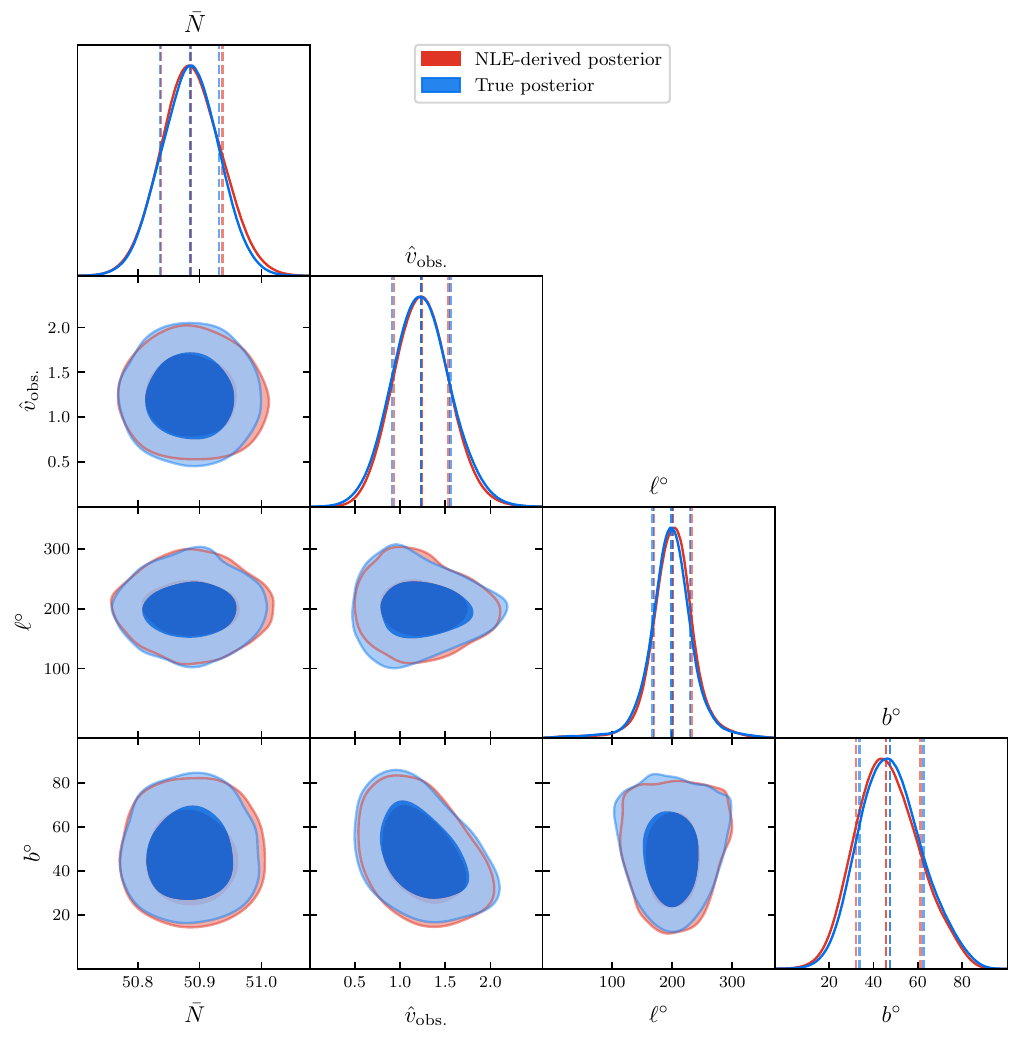}
  \end{center}
  \caption{
    Comparing the true posterior (blue) for a downscaled pseudo-CatSIM map
    and the NLE-inferred posterior (red).
    The results are virtually identical,
    verifying the accuracy of our pipeline in recovering dipole parameters.
    The details of the corner are the same as Fig.~\ref{fig:free_gauss_err_corner},
    except the credible interval titles on the 1D marginals have been supressed.
  }\label{fig:corner_evidence_acc}
\end{figure*}

We also need to check that the downscaling process is not interfering
with the Bayes factors: that is, the relative explanatory power (odds ratio)
of each model.
We take another pseudo-CatSIM map at $N_{\text{side}} = 64$
and introduce a dipole with $\hat{v}_{\text{obs.}} = 1.5$
($\mathcal{D} = 0.0075$)
and
$(l^\circ, b^\circ) = (230, 40)$.
Then we compute the true evidences for a model with a free dipole,
a dipole fixed to the CMB direction,
a dipole fixed to the CMB amplitude
and a dipole fixed to the CMB direction and amplitude.
We take the same map and downscale it to $N_{\text{side}} = 4$,
then use our NLE-inferred likelihood to determine the evidence for the same models.
After computing these, we write the Bayes factor for all models
with respect to the free dipole model.
These true Bayes factors are the blue points in Fig.~\ref{fig:lnB_scatter},
appearing at the left of each model column.
Meanwhile, the red points are the NLE-learned Bayes factors,
appearing at the right of each model column.
While the NLE-learned $\ln B$ estimates have significantly more scatter than the
true Bayes factors,
they are consistent given the error bars.
This, however, indicates that we have to be mindful of the uncertainty
behind our estimates for $\ln B$.

\section{SNLE vs SNPE results}
\label{sec:npe_verification}
\begin{figure*}
  \begin{center}
    \includegraphics[width=0.9\textwidth]{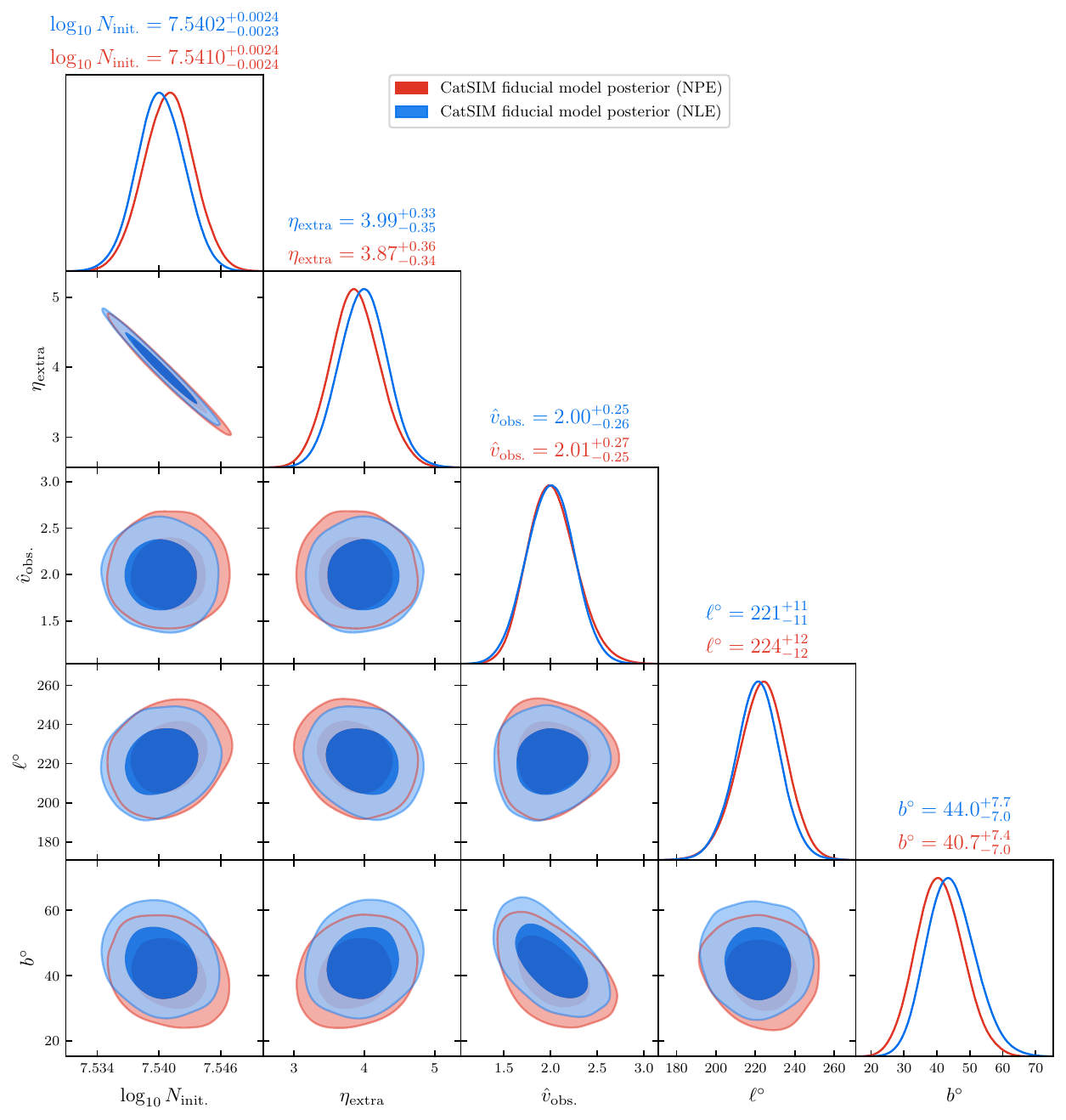}
  \end{center}
  \caption{
    The posterior distribution of our fiducial model
    `free dipole, extra error, Gaussian' derived using the NLE (blue) and NPE (red).
    The details of the corner are the same as Fig.~\ref{fig:free_gauss_err_corner},
    but we show two sets of credible intervals for the inferred parameters
    for each method (top/blue is the NLE, bottom/red is the NPE).
  }\label{fig:corner_npe_vs_nle}
\end{figure*}

We verify that our inferred posterior distribution for the fiducial model
(`free dipole, extra error, Gaussian')
is the same when using either the SNLE or SNPE.
Recall that Fig.~\ref{fig:free_gauss_err_corner} shows the posterior derived
from the SNLE.
We reproduce that same posterior in Fig.~\ref{fig:corner_npe_vs_nle}
but also overlay the results returned from the SNPE in red.
There are some very slight differences between the 1D marginals,
but the inferences made are virtually identical, especially looking at the
dipole parameters.
We stress that this illustrates the internal consistency of our SBI pipeline
since the SNPE and SNLE have highly disparate architectures (see Section~\ref{sec:appendix})
but yield the same conclusions.


\bsp	
\label{lastpage}
\end{document}

%% file: logb_summary.tex
\begin{tabular}{l r@{$\,\,\pm\,\,$}l}
\hline
Model & \multicolumn{2}{c}{$\ln B$} \\
\hline
Dipole from \citet{secrest2021} & $3.1$ & $1.2$ \\
Dipole from \citet{dam2023} & $2.4$ & $1.0$ \\
Free dipole, extra error, Gaussian & $0$ & $0$ \\
Free dipole, extra error, Student's $t$ & $-1.0$ & $0.9$ \\
CMB direction, free velocity & $-3.6$ & $0.9$ \\
CMB velocity, free direction & $-9.2$ & $1.1$ \\
CMB velocity \& direction & $-9.9$ & $1.2$ \\
Free dipole, no extra error, Student's $t$ & $-35.0$ & $1.3$ \\
Free dipole, no extra error, Gaussian & $-110.5$ & $5.2$ \\
\hline
\end{tabular}